\documentclass[aps,prb,twocolumn,showpacs,groupedaddress]{revtex4}
\usepackage{graphics,bm}
\usepackage{amssymb}
\usepackage{amsmath}
\usepackage{epsfig}
\usepackage{epsf}
\usepackage[usenames]{color}

\begin{document}

\title{Fluctuations of current-driven domain walls in the non-adiabatic regime.}

\author{M.E. Lucassen}
\email{m.e.lucassen@uu.nl}

\author{R.A. Duine}

\affiliation{Institute for Theoretical Physics, Utrecht
University, Leuvenlaan 4, 3584 CE Utrecht, The Netherlands}
\date{\today}

\begin{abstract}
We outline a general framework to determine the effect of
non-equilibrium fluctuations on driven collective coordinates, and
apply it to a current-driven domain wall in a nanocontact. In this
case the collective coordinates are the domain-wall position and
its chirality, that give rise to momentum transfer and spin
transfer, respectively. We determine the current-induced
fluctuations corresponding to these processes and show that at
small frequencies they can be incorporated by two separate
effective temperatures. As an application, the average time to
depin the domain wall is calculated and found to be lowered by
current-induced fluctuations. It is shown that current-induced
fluctuations play an important role for narrow domain walls,
especially at low temperatures.
\end{abstract}


\maketitle

\def\L{{\rm L}}
\def\R{{\rm R}}
\def\s{{\rm s}}
\def\p{{(+)}}
\def\m{{(-)}}
\def\K{{\rm K}}
\def\F{{\rm F}}
\def\kB{{\rm k}_{\rm B}}
\def\ii{{\rm i}}
\def\om{\mathbf{\Omega}}
\def\x{(x,t)}
\def\e{\varepsilon_{\rm F}}

\section{Introduction}

Fluctuations play an important role in many areas of physics. The
classic example is Brownian motion \cite{einstein1905}, for
example of a colloidal particle in a suspension. The effect of
collisions of the small particles, that constitute the suspension,
with the colloid is modelled by stochastic forces. The strength of
these forces is inferred from the famous fluctuation-dissipation
theorem, which states that their variance is proportional to
damping due to viscosity, and to temperature, and that their
average is zero. If the suspension is driven out of equilibrium,
the average force on the colloid will no longer be zero. Because
of the non-equilibrium situation, the fluctuation-dissipation
theorem in principle no longer holds, and the fluctuations cannot
be determined from it anymore. Another explicit example of
fluctuations in a driven system that do not obey the
fluctuation-dissipation theorem is shot noise in the current in
mesoscopic conductors, where the fluctuations are determined by
the applied voltage instead of temperature. It is ultimately
caused by the fact that the electric current is carried by
discrete charge quanta, the electrons.\cite{beenakker1997}

The non-equilibrium system on which we focus in this paper is a
current-driven domain wall \cite{berger1984,berger1985} in a
ferromagnetic conductor. Here the domain wall and the electrons
play the role of the colloid and the suspension from the above
example. There are two distinct processes that lead to
current-induced domain-wall motion: spin transfer
\cite{slonczewski1996,berger1996} and momentum
transfer.\cite{tatara2004} Physically, momentum transfer
corresponds to the force exerted on the domain wall by electrons
that are reflected by the domain wall or transmitted with
different momentum. Spin transfer corresponds to electrons whose
spin follows the magnetization of the domain wall adiabatically,
thereby exerting a torque on the domain wall. Most experiments
\cite{grollier2003,klaui2003,yamaguchi2004,beach2005,yamanouchi2006}
are in the adiabatic regime, where the electron spin follows the
direction of magnetization adiabatically and where the
spin-transfer torque is the dominant effect. The effect of spin
relaxation on spin transfer in the adiabatic limit, leading to a
dissipative spin-transfer torque, was discussed theoretically
\cite{zhang2004} and experimentally.\cite{hayashi2007,heyne2008}
The experiments by Feigenson {\it et al.} \cite{feigenson2007}
with SrRuO$_3$ films, on the other hand, are believed to be in the
non-adiabatic limit where domain walls are narrow compared to the
Fermi wave length and momentum transfer is dominant. In this
paper, we will mostly consider narrow domain walls in
nanocontacts.\cite{bruno1999,versluijs2001,garcia1999}

Apart from the forces and torques on the magnetization texture due
to nonzero average current, there are also current-induced
fluctuations on the magnetization \cite{foros2005,nunez2008} that
ultimately have their origin in shot noise in the spin and charge
current. Foros {\it et al.} \cite{foros2005} studied the effects
of spin-current shot noise in single-domain ferromagnets, and
found that for large voltage and low temperature the fluctuations
are determined by the voltage and not by the temperature.
Chudnovskiy {\it et al.} \cite{chudnovskiy2008} study spin-torque
shot noise in magnetic tunnel junctions, and in
Ref.~[\onlinecite{foros2008}] Foros {\it et al.} consider a
general magnetization texture and work out the current-induced
magnetization noise and inhomogeneous damping in the adiabatic
limit.

In this paper, we determine the effect of current-induced
fluctuations on a domain wall in the non-adiabatic limit. We show
that it leads to anisotropic damping and fluctuations and show
that the fluctuations can be described by two separate
voltage-dependent effective temperatures corresponding to momentum
transfer and spin transfer. We show that these effective
temperatures differ considerably from the actual temperature for
parameter values used in experiments with nanocontacts. From our
model, we also determine the momentum transfer and the adiabatic
spin-transfer torque on the driven domain wall, as well as the
damping corresponding to these processes.

\section{Model}

In this section, we present a model for treating a domain wall out
of equilibrium. We first develop a variational principle within
the Keldysh formalism, and then work out the various Green's
functions within Landauer-B\"uttiker transport.

\subsection{Keldysh Theory}

We consider a one-dimensional model of spins coupled to conduction
electrons. The action is on the Keldysh contour $C$ given by
\begin{align}\label{eq:hamiltonian}
S[\om,&\psi,\psi^*]=\int_Cdt\bigg\{-E_{\rm MM}[\om]
\nonumber\\
&+\int \frac{dx}{a}\bigg[-\hbar\mathbf{A}(\om\x)
\cdot\frac{\partial\om\x}{\partial t}\nonumber\\
&+\frac{\Delta}{2}\sum_{\sigma,\sigma'}
\psi^*_\sigma\x\mathbf{\Omega}\x
\cdot\boldsymbol{\tau}_{\sigma,\sigma'}
\psi_{\sigma'}\x\nonumber\\
&+\sum_\sigma\psi^*_\sigma\x \left(i\hbar\frac{\partial}{\partial
t }+\frac{\hbar^2\nabla^2}{2m}-V(x)\right)\psi_\sigma\x
\bigg]\bigg\}\;,\;
\end{align}
where $a$ is the lattice spacing, $\mathbf{A}(\om)$ is the
fictitious vector potential that obeys
$\boldsymbol{\nabla}_\om\times\mathbf{A(\om)}=\om$ and ensures
precessional motion of $\om$, $\Delta$ is the exchange-splitting
energy, $\om\x$ a unit vector in the direction of the
magnetization, $\boldsymbol{\tau}$ the vector of Pauli matrices,
and $V(x)$ an arbitrary scalar potential. The fields
$\psi^*_\sigma,\psi_\sigma$ represent the conduction electrons
with spin projection $\sigma\in\{\uparrow,\downarrow\}$. The
micromagnetic energy functional $E_{\rm MM}[\om]$ is given by
\begin{align}\label{eq:mmenergy}
E_{\rm MM}[\om]=
-\int\frac{dx}{a}&\Big[J\om\x\cdot\vec\nabla^2\om\x\nonumber\\
&-K_\perp\Omega_y^2\x+K_z\Omega_z^2\x\Big]\;,
\end{align}
with $J>0$ the spin stiffness and $K_\perp>0$ and $K_z>0$ the
hard- and easy-axis anisotropy constants, respectively. The
micromagnetic energy functional in Eq.~(\ref{eq:mmenergy}) has
stationary domain-wall solutions $\om(x)=(\sin\theta_{\rm
dw}\cos\phi_{\rm dw},\sin\theta_{\rm dw}\sin\phi_{\rm
dw},\cos\theta_{\rm dw})$.\cite{tatara2004} These stationary
solutions are the basis for a time-dependent variational {\it
ansatz} given by
\begin{align}
\theta_{\rm
dw}=2\arctan\left\{e^{[x-X(t)]/\lambda}\right\}\;;\qquad\phi_{\rm
dw}=\phi(t)\;,
\end{align}
where $\lambda$ is the domain-wall width. In the above, we have
taken the domain-wall position $X(t)$ to be time dependent.
Furthermore, $\phi(t)$ is the angle of the magnetization at the
center with the easy-plane, the so-called chirality. Using the
above {\it ansatz}, the first two terms in the action in
Eq.~(\ref{eq:hamiltonian}) simplify to
\begin{align}\label{eq:actionzero}
S_0[X,\phi]=\hbar N\int_C dt\left[\frac{\dot{X}}{\lambda}\phi
-\frac{K_\perp}{2}\sin2\phi\right]\;.
\end{align}
Here, $N=2\lambda/a$ is the number of spins in the domain wall.
Note that in three dimensions, the number of spins increases by a
factor $A/a^2$, where $A$ is the cross-sectional area of the
sample.

Stochastic forces are not obtained in a natural way by variation
of the real-time action or the Euclidean action of the system. The
functional Keldysh formalism,\cite{footnotestoof} however,
provides us with the (current-induced and thermal) noise terms
automatically, and is therefore more elegant for our purposes. By
doing perturbation theory in the collective coordinates $X$ and
$\phi$
\begin{align}\label{eq:perturbation}
\mathbf{\Omega}=\om\Big|_0 +X\frac{\partial\om}{\partial X}\Big|_0
+\phi\frac{\partial\om}{\partial \phi}\Big|_0 +{\rm h.o.}\;,
\end{align}
where from here onward the subscript $|_{_0}$ denotes evaluation
at $X=\phi=0$, we derive an effective action on the Keldysh
contour for the collective coordinates. We consider the
low-frequency limit, which is a good approximation because the
motion of the collective coordinates is on a much slower time
scale than the electronic system.

The total action is now given by
$S[X,\phi,\psi^*,\psi]=S_0[X,\phi]+S_{\rm
C}[X,\phi,\psi^*,\psi]+S_{\rm E}[\psi^*,\psi]$. The contribution
to the action in Eq.~(\ref{eq:hamiltonian}) that describes
coupling between magnetization and electrons is up to first order
given by
\begin{align}\label{eq:actioncoupling}
S_{\rm C}&[X,\phi]\Big|_0=\frac{\Delta}{2} \int_C dt\int
dx\sum_{\sigma,\sigma'}\psi^*_\sigma\x\nonumber\\
&\Big[(\partial_X\boldsymbol{\Omega}
\cdot\boldsymbol{\tau})\Big|_0X(t)
+(\partial_{\phi}\boldsymbol{\Omega}
\cdot\boldsymbol{\tau})\Big|_0\phi(t)\Big]\psi_{\sigma'}\x,
\end{align}
where $\sigma$ and $\sigma'$ denote the spin of the electrons. The
electron action reads
\begin{align}\label{eq:actionelectrons}
&S_{\rm E}[\psi^*,\psi]=\int_C dt\int
dx\sum_{\sigma,\sigma'}\; \psi^*_\sigma(x,t)\nonumber\\
&\left[\left(i\hbar\frac{\partial}{\partial
t}+\frac{\hbar^2\nabla^2}{2m} -V(x)\right)\delta_{\sigma\sigma'}
-V_{\sigma\sigma'}(x)\right]\psi_{\sigma'}(x,t)\;,
\end{align}
with the potential $V_{\sigma\sigma'}$, which arises from the
zeroth order term $\om|_0$, given in Eq.~(\ref{eq:potential})
below. The perturbation theory in $X$ and $\phi$ enables us to
derive an effective action on the Keldysh contour for these
coordinates
\begin{align}\label{eq:actioneffective}
S_{\rm eff}&[X,\phi]\simeq S_0[X,\phi]+\langle S_{\rm
C}[X,\phi,\psi^*,\psi]\rangle\nonumber\\
+&\frac{i}{2\hbar}\Big(\langle S_{\rm
C}^2[X,\phi,\psi^*,\psi]\rangle-\langle S_{\rm
C}[X,\phi,\psi^*,\psi]\rangle^2 \Big)\;.
\end{align}
Here, the expectation values are taken with respect to the
electron action $S_{\rm E}[\psi^*,\psi]$ in
Eq.~(\ref{eq:actionelectrons}), i.e.,
\begin{align}\langle
O[X,\phi,&\psi^*,\psi]\rangle=\nonumber\\
&\int d[\psi^*]d[\psi]e^{iS_{\rm
E}[\psi^*,\psi]/\hbar}O[X,\phi,\psi^*,\psi]\;.
\end{align}
In the next section, we evaluate these expectation values in more
detail.

Since we now have an effective action as a function of the
collective coordinates $X$ and $\phi$, we can make use of the
advantages of the Keldysh formalism. The effective action in
Eq.~(\ref{eq:actioneffective}) is integrated from $t=-\infty$ to
$t=\infty$ and back.
\begin{figure}
\includegraphics[width=8cm]{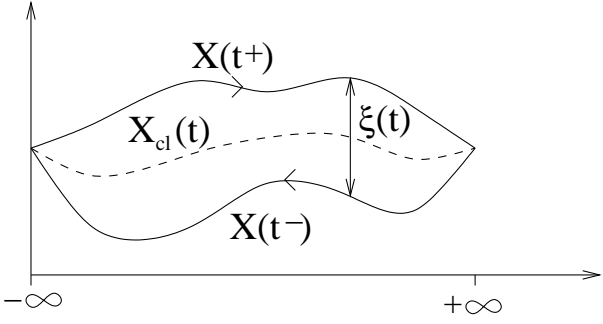}
\caption{A path that the coordinate $X$ might take on the Keldysh
contour. The deviation from the (classical) mean path is denoted
by $\xi(t)$.}\label{fig:keldyshcontour}
\end{figure}
The forward and backward paths are different, as is shown for the
coordinate $X$ in Fig.\ref{fig:keldyshcontour}, such that we write
\begin{align}
 X(t^\pm)=X_{\rm cl}(t)\pm \frac{\xi(t)}{2}\;;\qquad
&\phi(t^\pm)=\phi_{\rm cl}(t)\pm
\frac{\kappa(t)}{2}\;;\nonumber\\
\int_{C}dt f(t)
=\int_{-\infty}^{\infty}dt^+f(t^+)&+\int_{\infty}^{-\infty}dt^-f(t^-)\;,
\end{align}
with the assumption that the variations $\xi$ and $\kappa$ are
small. Furthermore, they obey the boundary conditions
$\xi(\pm\infty)=\kappa(\pm\infty)=0$. Integrating the effective
action over this contour and using the method outlined in
Refs.~[\onlinecite{footnotestoof,nunez2008}], we ultimately obtain
the Langevin equations for a domain wall
\begin{align}\label{eq:langevinX}
\frac{\dot{X}_{\rm cl}(t)}{\lambda}-\alpha_{\phi}\dot{\phi}_{\rm
cl}(t) =K_\perp\cos2\phi_{\rm cl }(t)+F_{\phi}+&\eta_{\phi}(t)\;,
\\
\label{eq:langevinphi} \dot{\phi}_{\rm cl
}(t)+\alpha_X\frac{\dot{X}_{\rm cl}(t)}{\lambda} =-F_X+\eta_X(t)
\;.
\end{align}
The stochastic contributions $\eta_{\phi}$ and $\eta_X$ in this
expression arise via a Hubbard-Stratonovich transformation of
terms quadratic in $\xi$ and $\kappa$.

The expectation value of the action $S_C[X,\phi,\psi^*,\psi]$ in
the effective action in Eq.~(\ref{eq:actioneffective}) provide us
with the forces
\begin{align}\label{eq:force}
F_i=\frac{\Delta}{\hbar
N}\sum_{\nu,\sigma,\sigma'}G^<_{\nu,\sigma,\sigma'}(0)
\langle\nu,\sigma|\partial_i\boldsymbol{\Omega}|_0
\cdot\boldsymbol{\tau}|\nu,\sigma'\rangle\;,
\end{align}
where the index $i\in\{X/\lambda,\phi\}$. Note that in this
expression, $F_{\phi}$ corresponds to spin transfer and $F_X$ to
momentum transfer. As we will see later on,
$\partial_{\phi}\mathbf{\Omega}|_0\cdot\boldsymbol{\tau}$ is
associated with the divergence of the spin current, and
$\partial_{X}\mathbf{\Omega}|_0\cdot\boldsymbol{\tau}$ is
associated with the force of the domain wall on the conduction
electrons. The latter is, in the absence of disorder, proportional
to the resistance of the domain wall.\cite{tatara2004} The lesser
Green's function in this expression is defined by
$G_{\nu,\sigma,\sigma'}(t,t')=
\theta(t,t')G^>_{\nu,\sigma,\sigma'}(t-t')
+\theta(t',t)G^<_{\nu,\sigma,\sigma'}(t-t')$, where the Heaviside
step functions are defined on the Keldysh contour. Note that the
lesser Green's function in Eq.~(\ref{eq:force}) is evaluated at
equal times $t=t'$. Furthermore, we have expanded the Keldysh
Green's function according to $\ii
G_{\sigma,\sigma'}(x,t;x',t')\equiv \langle\psi_\sigma(x,t)
\psi^*_{\sigma'}(x',t')\rangle=\ii\sum_\nu
G_{\nu,\sigma,\sigma'}(t,t')\chi_{\nu,\sigma}(x)
\chi_{\nu,\sigma'}^*(x')$, where $\chi$ and $\chi^*$ are electron
eigenstates in the presence of a static domain wall that are
labeled by $\nu$. In terms of these states, the matrix elements
are defined as
\begin{align}
\int dx\; \langle\nu,\sigma|\hat{O}|\nu',\sigma'\rangle=\int
dx\chi_{\nu,\sigma}(x)\hat{O}_{\sigma,\sigma'}(x)\chi_{\nu',\sigma'}^*(x)
\;.
\end{align}

The damping terms in
Eqs.~(\ref{eq:langevinX})~and~(\ref{eq:langevinphi}) follow from
the second-order terms in the perturbation theory in $X$ and
$\phi$ and read $\alpha_i=\mp {\rm
Im}[\tilde\Pi^{(\pm)}_i(\omega)]/(N\hbar \omega)$ for
$\omega\rightarrow 0$, with $\tilde\Pi^{(\pm)}(\omega)$ the
response function given below. Since the action in
Eq.~(\ref{eq:hamiltonian}) is quadratic in the electron fields, we
use Wick's theorem to write the response function in terms of
electron Green's functions
\begin{align}\label{eq:damping}
&\tilde\Pi^{(\pm)}_i(\omega)=\mp\frac{\Delta^2}{4\hbar}\sum_{\nu,\nu'}
\sum_{\sigma,\sigma'}\sum_{\rho,\rho'}\nonumber\\
&\times\Big[G^>_{\nu,\sigma,\rho'}(\omega)
G^<_{\nu',\rho,\sigma'}(-\omega)-G^<_{\nu,\sigma,\rho'}(\omega)
G^>_{\nu',\rho,\sigma'}(-\omega)\Big]\nonumber\\
&\times \langle\nu,\sigma|\partial_i\boldsymbol{\Omega}|_0
\cdot\boldsymbol{\tau}|\nu',\sigma'\rangle
\langle\nu',\rho|\partial_i\boldsymbol{\Omega}|_0
\cdot\boldsymbol{\tau}|\nu,\rho'\rangle\;.
\end{align}
The functions $G^>(\pm\omega)$ and $G^<(\pm\omega)$ denote Fourier
transforms of $G^>(\pm(t-t'))$ and $G^<(\pm(t-t'))$, respectively.

Without needing to assume (approximate) equilibrium, the Keldysh
formalism provides us with an expression for the strength of the
fluctuations in both coordinates
$\langle\eta_i(t)\eta_{j}(t')\rangle
=\ii\delta_{i,j}\Pi^\K_i(t-t')/(\hbar N^2)\simeq
\ii\delta_{i,j}\tilde{\Pi}^\K_i(\omega=0)\delta(t-t')/(\hbar N^2)$
and $\langle\eta_i(t)\rangle=0$. The Keldysh component of the
response function contains similar matrix elements as the damping
terms and is given by
\begin{align}\label{eq:stochasticstrength}
&\tilde\Pi^\K_i(\omega)= \frac{\Delta^2}{2\hbar}\sum_{\nu,\nu'}
\sum_{\sigma,\sigma'}\sum_{\rho,\rho'}\nonumber\\
&\times\Big[G^>_{\nu,\sigma,\rho'}(\omega)
G^<_{\nu',\rho,\sigma'}(-\omega)+G^<_{\nu,\sigma,\rho'}(\omega)
G^>_{\nu',\rho,\sigma'}(-\omega)\Big]
\nonumber\\
&\times\langle\nu,\sigma|\partial_i\boldsymbol{\Omega}|_0
\cdot\boldsymbol{\tau}|\nu',\sigma'\rangle
\langle\nu',\rho|\partial_i\boldsymbol{\Omega}|_0
\cdot\boldsymbol{\tau}|\nu,\rho'\rangle\;.
\end{align}
We now define two separate effective temperatures
\begin{align}\label{eq:efftemperature}
\kB T_{{\rm
eff},i}\equiv\frac{\ii\tilde\Pi_i^\K(\omega=0)}{2\alpha_i N^2}\;.
\end{align}
These effective temperatures are defined such that
Eqs.~(\ref{eq:langevinX})~and~(\ref{eq:langevinphi}) obey the
fluctuation-dissipation theorem with the effective temperatures.
In the absence of a bias voltage, the effective temperatures
reduce to the actual temperature divided by the number of spins in
the system. More general, the effective temperatures are
proportional to $1/N$, which is understood because they describe
fluctuations in collective coordinates made up of $N$ degrees of
freedom.\cite{duine2007} We note that our formalism applies to any
set of collective coordinates, and is not necessarily restricted
to the example of a domain wall. We also point out that going
beyond the low-frequency limit and taking into account the full
frequency dependence in Eq.~(\ref{eq:stochasticstrength}) leads to
colored noise. In this case, effective temperatures may no longer
be unambiguously defined.\cite{mitra2005}

\subsection{Landauer-B\"{u}ttiker transport}

We now evaluate Eqs.~(\ref{eq:force}--\ref{eq:stochasticstrength})
using the Landauer-B\"{u}ttiker formalism, {\it i.e.}, the
scattering theory of electronic transport. In order for this
formalism to apply, the phase-coherence length $L_\phi$ must be
larger than the domain-wall width $\lambda$. To compute the terms
in the Langevin equations
(\ref{eq:langevinX})~and~(\ref{eq:langevinphi}) explicitly, we
need to find the matrix elements
$\langle\nu,\sigma|\partial_i\mathbf{\Omega}|_0\cdot
\boldsymbol{\tau}|\nu',\sigma'\rangle$ and the Green's functions.
The Keldysh Green's function is in terms of scattering states
$\chi^{\zeta\kappa\varepsilon}_{\sigma}(x)$ given by
\begin{align}\label{eq:greensfunction}
&\ii G_{\sigma,\sigma'}(x,t;x',t')\qquad\qquad\qquad\nonumber\\
&=\int_{0}^{\infty} \frac{d\varepsilon}{2\pi}
\sum_{\zeta,\kappa}\frac{2m/\hbar^2} {k_{\zeta\kappa}}
e^{-\frac{\ii}{\hbar} \varepsilon(t-t')}
\chi^{\zeta\kappa\varepsilon}_{\sigma'}(x)
[\chi^{\zeta\kappa\varepsilon}_\sigma(x')]^*\nonumber\\
&\times\{\theta(t,t')[1-N_{\rm
F}(\varepsilon-\mu_\zeta)]-\theta(t',t)N_{\rm
F}(\varepsilon-\mu_\zeta)\}\;,
\end{align}
with $\mu_\zeta$ the chemical potential of the lead on side
$\zeta\in\{{\rm L(eft),R(ight)}\}$,
$\kappa\in\{\uparrow,\downarrow\}$ the spin of the incoming
particles and $N_{\rm F}(x)$ the Fermi distribution function. We
choose $V(x)=0$ for convenience. The momenta $k_{\zeta\kappa}$
associated with an energy $\varepsilon$ are given by
$k_{\L\uparrow}=k_{\R\downarrow}
=k_\F\sqrt{(\varepsilon+\Delta/2)/\e\;}$ and
$k_{\R\uparrow}=k_{\L\downarrow}
=k_\F\sqrt{(\varepsilon-\Delta/2)/\e\;}$, where
$k_\F=\sqrt{2m\e/\hbar^2}$, with $\e$ the Fermi energy in the
leads. Note that the index $\nu$ used earlier now contains
information on the origin, spin and energy of the incoming
particle. We define the asymptotic expression for the scattering
states in terms of transmission and reflection coefficients,
\begin{align}\label{eq:scatteringstates}
\chi_\sigma^{\L\kappa\varepsilon}\!=\!\left\{\begin{array}{cl}
\!\!\delta_{\sigma,\kappa}e^{\ii k_{\L\kappa}x}\! +
\delta_{\sigma,\gamma}\sqrt{\frac{k_{\L\kappa}}{k_{\L\gamma}}}r_{\gamma\kappa}
(\varepsilon)e^{-\ii k_{\L\gamma}x},
&x\rightarrow\!-\infty\;;\\\\
\delta_{\sigma,\gamma}\sqrt{\frac{k_{\L\kappa}}{k_{\R\gamma}}}t_{\gamma\kappa}
(\varepsilon)e^{\ii k_{\R\gamma}x},
&x\rightarrow\!+\infty\;,\end{array}\right.\nonumber\\
\end{align}
where summation over spin-index $\gamma\in\{\uparrow,\downarrow\}$
is implied, and with a similar expression for right-incoming
particles. From the explicit form of the {\it ansatz}, it is
easily seen that $\partial\om/\partial X=-\partial\om/\partial x$
which enables us to write $(\Delta/2)(\partial\om/\partial
X)|_0\cdot\boldsymbol{\tau}_{\sigma,\sigma'}=\partial_xV_{\sigma,\sigma'}(x)$.
Here $\partial_x$ denotes a derivative with respect to $x$, and
the potential is given by
\begin{align}\label{eq:potential}
V_{\sigma\sigma'}(x)=
-\frac{\Delta}{2}\begin{pmatrix}\cos\theta_{\rm dw}&\sin\theta_{\rm dw}\\
\\ \sin\theta_{\rm dw}&-\cos\theta_{\rm dw}\end{pmatrix}\Bigg|_0\;.
\end{align}
Furthermore, one can check that
$(\partial\om/\partial\phi)\cdot\boldsymbol{\tau}
_{\sigma\sigma'}|_0
=-(\boldsymbol{\tau}_{\sigma\sigma'}\times\om)|^z_0$, where
uppercase $z$ denotes the $z$ component of this cross product. The
expectation value of this quantity is directly related to the
divergence of the spin-current $J_{\rm s}^z$, which measures the
$z$ component of the total spin-current
\begin{align}\label{eq:spinpart}
\partial_xJ^z_{\rm s}(x)
=\frac{\Delta}{2}\sum_{\sigma,\sigma'}\chi^*_\sigma(x)
\Big[\boldsymbol{\tau}_{\sigma\sigma'}
\times\mathbf{\Omega}(x)\Big]^z_0\chi_{\sigma'}(x)\;.
\end{align}
In this expression, $\chi$ is a solution to the zeroth order
time-independent Schr\"odinger equation with the potential
$V_{\sigma,\sigma'}(x)$. The spin current is defined as
\begin{align}\label{eq:spincurrent}
J_{\rm s}^z(x)=\frac{\hbar^2}{4mi}\sum_{\sigma,\sigma'}\Big\{&
\chi^*_\sigma(x)\tau_{\sigma\sigma'}^z[\partial_x\chi_{\sigma'}(x)]\nonumber\\
&\;\;-[\partial_x\chi^*_\sigma(x)]\tau_{\sigma\sigma'}^z
\chi_{\sigma'}(x)\Big\}\;.
\end{align}
From this, we observe that $F_\phi$ determined by
Eq.~(\ref{eq:force}) is indeed proportional to the divergence of
spin current and hence corresponds to spin transfer.

We define $\mu_\L=\mu+|e|V$ and $\mu_\R=\mu\simeq\e$. The
expressions for the Green's function and the scattering states in
Eqs.~(\ref{eq:greensfunction})~and~(\ref{eq:scatteringstates}) now
allow us to write Eqs.~(\ref{eq:force}--\ref{eq:efftemperature})
in terms of transmission and reflection coefficients, the applied
voltage $V$, and $k_{\rm F}\lambda$. For example, the momentum
transfer in Eq.~(\ref{eq:force}), with $i=X$, is up to first order
in $|e|V/\e$ given by
\begin{align}\label{eq:momentumtransfer}
&F_X\simeq \frac{|e|V}{2\pi\hbar
N}k_{\rm F}\lambda\nonumber\\
&\Bigg[\sqrt{\frac{\varepsilon+\Delta/2}{\e}}\Big(1+R_{\uparrow\uparrow}-T_{\downarrow\downarrow}
+R_{\uparrow\downarrow}-T_{\downarrow\uparrow}\Big)\nonumber\\
&+\sqrt{\frac{\varepsilon-\Delta/2}{\e}}\Big(1+R_{\downarrow\downarrow}-T_{\uparrow\uparrow}
+R_{\downarrow\uparrow}-T_{\uparrow\downarrow}\Big) \Bigg]\;,
\end{align}
Here, the reflection and transmission coefficients are defined as
$R_{\sigma\sigma'}=R_{\sigma\sigma'}(\varepsilon=\e)$ with
$R_{\sigma\sigma'}(\varepsilon)\equiv|r_{\sigma\sigma'}(\varepsilon)|^2$,
and equivalently for the transmission coefficients. Note that,
although the coefficients are evaluated at the Fermi energy, they
also depend on the ratio $\Delta/\e$. The expression for the
momentum transfer in Eq.~(\ref{eq:momentumtransfer}) clearly
demonstrates its correspondence to electrons scattering off the
domain wall: it increases for increasing reflection and decreases
for increasing transmission.

The explicit form of the spin-transfer torque $F_\phi$ has
$\propto T_{\downarrow\uparrow}$ as leading term, which is a
measure for the number of electrons that follow the domain-wall
magnetization.

\begin{figure}[h!]
\includegraphics[width=8.5cm]{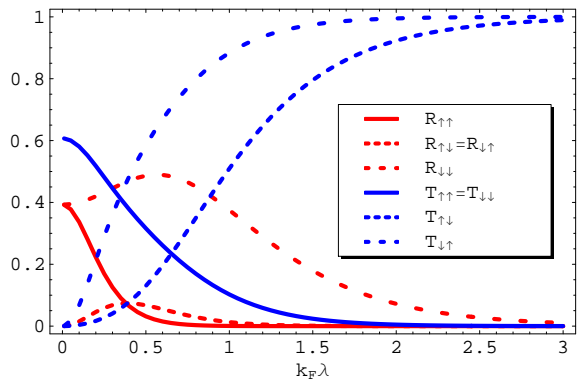}
\caption{(color online). Transmission and reflection coefficients
as functions of $k_{\rm F}\lambda$ for $\Delta/2\e=0.8$ and
$V=0$.}\label{fig:coefficients}
\end{figure}

The reflection and transmission coefficients are obtained by
solving the Schr\"{o}dinger equation of the system numerically,
and matching the results to the asymptotic behavior in
Eq.~(\ref{eq:scatteringstates}). As an example, we present the
coefficients for $\Delta/2\e=0.8$ as a function of $k_{\rm
F}\lambda$ in Fig.~\ref{fig:coefficients}.

\section{Results}

As we have shown in the previous section, we are able to express
Eqs.~(\ref{eq:force}--\ref{eq:stochasticstrength}) in terms of
transmission and reflection coefficients using Landauer-B\"uttiker
transport, like in Eq.~(\ref{eq:momentumtransfer}). As indicated,
these coefficients are obtained by numerically solving the
Schr\"odinger equation.

\begin{figure}[h!]
\includegraphics[width=8.5cm]{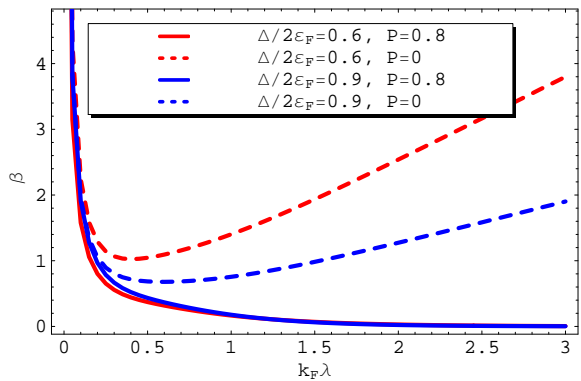}
\caption{(color online). The parameter $\beta=F_X/F_{\phi}$ as a
function of $k_{\rm F}\lambda$ for $\Delta/2\e=0.6$ and
$\Delta/2\e=0.9$, both at zero voltage. The dashed curves are
obtained by ignoring polarization, the solid lines are obtained
for a polarization of $P=0.8$.}\label{fig:beta}
\end{figure}

In the limit of vanishing voltage, Eq.~(\ref{eq:momentumtransfer})
is the exact expression for the momentum transfer. In fact, the
momentum transfer as well as the spin-transfer torque are for
small $|e|V/\e$ proportional to the voltage, in agreement with the
fact that these quantities are usually described as linear with
the spin current.\cite{tatara2004} We present the ratio of these
forces that measures the degree of nonadiabaticity, denoted by
$F_X/F_{\phi}=\beta$, as a function of $k_\F\lambda$ for several
values $\Delta/2\e$ in Fig.~\ref{fig:beta}.

In Fig.~\ref{fig:beta}, the dashed curves show the result obtained
directly from Eq.~(\ref{eq:momentumtransfer}) and an equivalent
expression for $F_{\phi}$. We see that $\beta$ is large for small
$k_{\rm F}\lambda$, as expected. The ratio, however, does not
vanish for large $k_{\rm F}\lambda$, which one would expect, but
instead acquires a linear dependence on $k_{\rm F}\lambda$.
Mathematically, this is caused not by an increase of momentum
transfer, but instead by a vanishing spin transfer. We can make
sure that the spin transfer does not vanish by taking into account
the polarization of the incoming electron
current.\cite{mazin1999,waintal2004}. If we do take this into
account, such that $R_{\sigma\uparrow}\rightarrow
R_{\sigma\uparrow}(1+P)$, $R_{\sigma\downarrow}\rightarrow
R_{\sigma\downarrow}(1-P)$, and equivalently for transmission
coefficients, we find for $P=0.8$ the solid curves in
Fig.~\ref{fig:beta}. We see that these curves indeed go to zero in
the adiabatic limit $k_{\rm F}\lambda\gg1$. From
Fig.~\ref{fig:beta}, it is clear that the polarization plays a big
role from values $k_\F\lambda\simeq0.3$ onwards. Note that our
theory does not take into account the dissipative spin-transfer
torque, which gives similar contributions as momentum
trensfer.\cite{zhang2004}

\begin{figure}[h!]
\includegraphics[width=8.5cm]{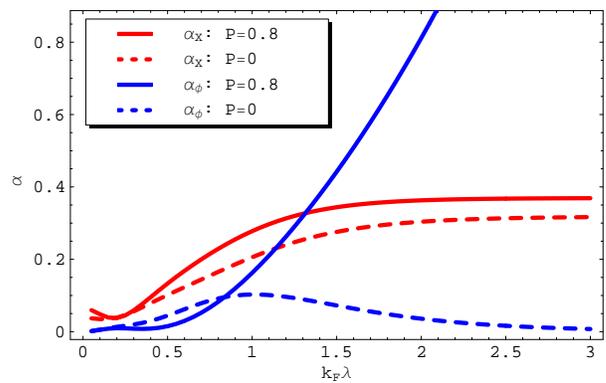}
\caption{(color online). The damping parameters $\alpha_X$ and
$\alpha_{\phi}$ as a function of $k_{\rm F}\lambda$ for
$\Delta/2\e=0.9$, both at zero voltage. The dashed curves are
obtained by ignoring polarization, the solid lines are obtained
for a polarization of $P=0.8$.}\label{fig:damping}
\end{figure}

For the damping parameters $\alpha_X$ and $\alpha_{\phi}$, we find
that for small $|e|V/\e$, they both acquire corrections linear in
the voltage, in agreement with Katsura {\it et al.}
\cite{katsura2006} and N\'u\~nez and Duine.\cite{nunez2008} The
dependence on $k_\F\lambda$ is much less trivial, as is shown in
Fig.~\ref{fig:damping}, where the curves are taken at zero
voltage.

The unpredictable behavior of the damping parameters as a function
of $k_\F\lambda$ for small $k_\F\lambda$ arises from the details
of the solutions of the Schr\"odinger equation. For large
$k_\F\lambda$, we see that without polarization $\alpha_\phi$ goes
to zero, whereas for nonzero polarization, it increases
quadratically. This is understood from the fact that damping in
the angle $\phi$ arises from emission of spin waves. This in its
turn is closely related to spin-transfer torque, which goes to
zero for $P=0$ but assumes nonzero values for $P>0$, as was
discussed earlier in this section. It should be noted, however,
that this approach breaks down for large values $k_\F\lambda$
since we then lose phase coherence as $\lambda>L_\phi$.
Furthermore, the polarization could be addressed in a more
rigorous way by taking into account more transverse channels. Note
that the fact that $\alpha_X\neq\alpha_{\phi}$ is a specific
example of inhomogeneous damping as discussed by Foros {\it et
al.}.\cite{foros2008}

\begin{figure}[h!]
\includegraphics[width=8.5cm]{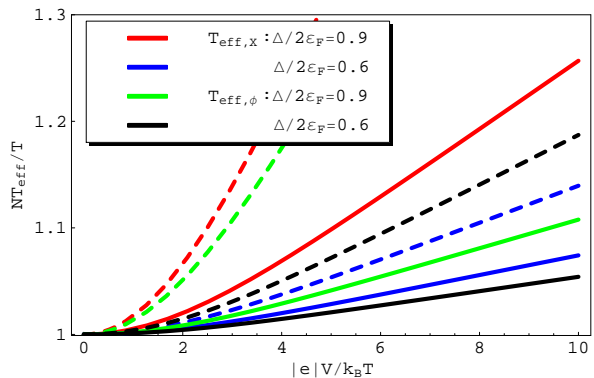}
\caption{(color online). The effective temperatures $NT_{{\rm
eff},X}/T$ and $NT_{{\rm eff},\phi}/T$ as a function of $|e|V/\kB
T$ for $\Delta/2\e=0.6$ or $\Delta/2\e=0.9$, all at
$k_\F\lambda=1$. The dashed curves are for $P=0$, the solid curves
are for $P=0.6$.}\label{fig:teff}
\end{figure}

The effective temperatures of the system depend on the
dimensionless parameters $k_\F\lambda$ and $\Delta/2\e$. In
Fig.~\ref{fig:teff} we plot the effective temperatures for $X$ and
$\phi$ as functions of $|e|V/\kB T$ for $k_\F\lambda=1$ and
several values $\Delta/2\e$.
The solid curves are obtained for $P=0.6$, the dashed curves do
not take into account polarization. Note that the effective
temperatures due to current-induced fluctuations can be
substantially larger than the actual temperature, and are for
large voltage proportional to $|e|V$.

\begin{figure}[h!]
\includegraphics[width=8.5cm]{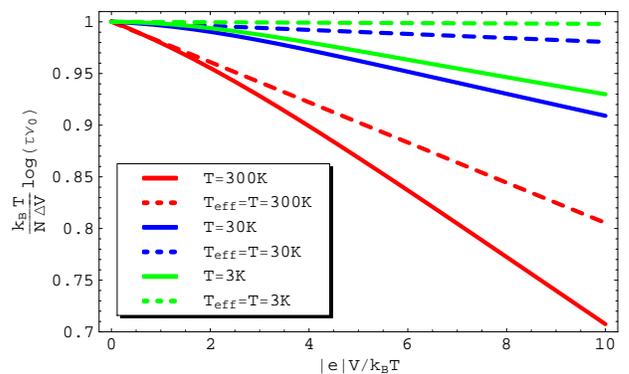}
\caption{(color online). Logarithm of the escape time $(\kB
T/N\Delta V) \log{\tau\nu_0}$ as a function of $|e|V/\kB T$ with
(solid curves) and without (dashed curves) current-induced
fluctuations. We used $\Delta V\lambda/\xi=1$ meV, $N=10$,
$k_\F\lambda=1$, $\Delta=\e$, $P=0$, and several
temperatures.}\label{fig:logtau}
\end{figure}

As an application of the effective temperatures derived above, we
compute depinning times as a function of the voltage. The
effective temperature $T_{\rm eff,X}$ influences depinning from a
spatial potential for the domain wall, such as a nanoconstriction.
We model the pinning potential by a potential well of width
$2\xi$, given by $V=\Delta V(X^2/\xi^2-1)\theta(|X|-\xi)$
\cite{tatara2004}. Using Arrhenius' law, the escape time is given
by $\log(\tau\nu_0)=N(\Delta V-\hbar F_X\xi/\lambda)/\kB T_{\rm
eff,X}$, where $\nu_0\sim\Delta V/\hbar$ is the attempt frequency.
Note that the effect of the momentum transfer is determined by the
ratio $\xi/\Delta V$, and that the force itself is still dependent
on the number of spins $N$. We show our results for $\Delta
V\lambda/\xi=1$ meV, $N=10$ and several temperatures in
Fig.~\ref{fig:logtau}. The results for $T_{\rm eff}=T$ are also
shown. Note that current-induced fluctuations decrease depinning
times with respect to the result with the actual temperature.
Here, we did not take into account polarization, which reduces
this effect since polarization brings down the effective
temperature, as was seen in Fig.~\ref{fig:teff}.

\section{Discussion}

We have established a microscopic theory that describes the
effects of current-induced fluctuations on a domain wall. Since
fluctuations in the current influence the system via spin transfer
and momentum transfer, we find two separate forces, dampings and
effective temperatures that correspond to these processes. We note
that the ratio of the momentum transfer and the spin transfer
$\beta=F_X/F_{\phi}$ that we calculate does not yet include the
contribution due to spin relaxation. However, this contribution is
small compared to the contribution due to momentum transfer when
the domain wall is narrow, and can therefore be ignored. In
addition to the contribution due to the coupling of the domain
wall with the electrons in the leads, that we consider here, there
is an intrinsic contribution to the damping due to
spin-relaxation, which is of the order $\alpha_0\sim0.01-0.1$ in
bulk materials, {\it i.e.}, of the same order as the
voltage-dependent damping parameters that we obtain. A
voltage-independent contribution to the damping will decrease the
effective temperature, and thereby increase the depinning time
somewhat.

As an application, we have studied depinning of the domain wall
from a nanoconstriction. The width of the domain wall in
nanocontacts is approximately the same as the nanocontact itself
\cite{bruno1999}. In experiments, it can be as small as
$\lambda\sim 1$ nm \cite{versluijs2001,garcia1999}, which is
smaller than the phase coherence length $L_\phi\sim 10$ nm in
metals at room temperature and therefore permits a
Landauer-B\"{u}ttiker transport approach. Tatara {\it et al.}
\cite{tatara1999} have shown that in nanocontacts in metals Ni and
Co, the exchange-splitting energy can reach high values
$\Delta/2\e\simeq 0.98$. The voltage on the system in the
experiment by Coey {\it et al.} is of the order $|e|V\sim 0.1$ eV,
which leads to $|e|V/\kB T\simeq 4$ at room temperature. The
potential barrier in experiments on nanocontacts for typical
displacements $\xi\sim10\lambda$ \cite{versluijs2001} is very
large $\Delta V\sim 10$ eV, but can be tuned by applying an
external magnetic field. We see from Fig.~\ref{fig:logtau} that at
room temperature, the current-induced fluctuations already have an
effect on depinning times, even if we take into account the fact
that polarization might reduce this effect somewhat. At lower
temperatures, this effect becomes larger. Under these
circumstances, Coey {\it et al.} find no evidence for heating
effects, which would be another source of increased fluctuations.
Therefore, current-induced fluctuations should be observable with
domain walls in nanocontacts.

Depinning of the angle $\phi$ is possible for relatively low
values of the transverse anisotropy $K_\perp$. This depinning
corresponds to switching between N\'{e}el walls of different
chirality. Between the N\'{e}el wall configurations, the domain
wall takes the form of a Bloch wall, that has higher energy. Coey
{\it et al.} \cite{coey2001} have argued that in
nanoconstrictions, the energy difference is comparable to the
thermal energy at room temperature. Now, $T_{{\rm eff},\phi}$ is
the effective temperature of interest, and from
Fig.~\ref{fig:teff} we observe that current-induced fluctuations
substantially alter this temperature. We therefore expect that
effects of current-induced fluctuations on fluctuation-assisted
domain-wall transformations can be significant.

This work was supported by the Netherlands Organization for
Scientific Research (NWO) and by the European Research Council
(ERC) under the Seventh Framework Program. It is a pleasure to
thank Henk Stoof for discussions.

\end{document}